\documentclass[reprint,prx,showpacs,superscriptaddress]{revtex4-1}
\usepackage{amsmath,amssymb,times,color,graphicx,multirow}
\usepackage[colorlinks=true,citecolor=blue,linkcolor=blue]{hyperref}
\usepackage{calligra}
\begin{document} 

\title{Unified understanding of symmetry indicators for all internal symmetry classes}

\author{Seishiro Ono}
\affiliation{The Institute for Solid State Physics, University of Tokyo, Kashiwa 277-8581, Japan}

\author{Haruki Watanabe}  
\email{haruki.watanabe@ap.t.u-tokyo.ac.jp}
\affiliation{Department of Applied Physics, University of Tokyo, Tokyo 113-8656, Japan}

\begin{abstract}
The interplay between symmetry and topology in electronic band structures has been one of the central subjects in condensed-matter physics.  Recently, it has been getting clear that 
a wide variety of useful information about the band topology can be extracted by focusing the symmetry representations of valence bands without computing Wilson loops.
In this work, we extend the previous studies on this subject to all 10 Altland-Zirnbauer symmetry classes in each of 230 space groups. We derive various general statements that should be useful in the search for topological superconductors and topological semimetals.
\end{abstract}

\maketitle

\section{Introduction}
The discoveries of topological insulators, topological superconductors~\cite{RevModPhys.82.3045,RevModPhys.83.1057}, and topological semimetals~\cite{RevModPhys.90.015001} have attracted researchers around the world because of their robust surface properties and intriguing bulk responses, which will have many future applications in the innovation of new devices.  Establishing the full classification of all possible topologically distinct band structures, given the specific symmetry setting and the dimensionality of the system, is one of the ultimate goals of theoretical studies of topological phases.  The ``topological periodic table"~\cite{PhysRevB.78.195125,doi:10.1063/1.3149495,ryu2010topological} is a milestone in this line of research, providing the complete classification of topological insulators for each of the 10 Altland-Zirnbauer (AZ) symmetry classes~\cite{PhysRevB.55.1142} in each spatial dimension.  The AZ symmetry class concerns the combination of thee internal symmetries -- the time-reversal symmetry (TRS) $\Theta$, the particle-hole symmetry (PHS) $\Xi$, and the chiral symmetry (CS) $\Pi$ -- which cannot be simply diagonalized together with the Hamiltonian.

Apart from these internal symmetries and the lattice translation symmetry, crystalline materials may also posses spatial symmetries such as the spatial inversion, mirror refections, and $n$-fold rotations as well as other nonsymmorphic symmetries. The combination of these operations give rise to in total 230 different space groups in three dimensions and 80 distinct layer groups~\cite{ITC}. The latter is relevant for two dimensional systems embedded in three dimensions like graphene. The spatial symmetry can also protect new topological phases, leading to the notion of topological crystalline insulators.  Mirror Chern insulators protected by a mirror symmetry~\cite{PhysRevLett.106.106802} are perhaps the most well-established example, but higher-order topological insulators (HOTIs)~\cite{Schindlereaat0346,PhysRevB.97.205136,PhysRevLett.119.246401,PhysRevB.97.205135} that exhibit lower dimensional gapless modes on the surfaces are also stabilized by crystalline symmetries.  There are many on-going attempts \cite{Robert_requested,Freed2013,Read,PhysRevX.7.041069,Shiozaki} trying to establish the complete characterization of topological insulators for every combination of AZ symmetry classes and spacial symmetries. 

Not only giving birth to new topological phases, the spatial symmetries also play an important role in the diagnosis of the topological property of band structures. The representative example of this phenomenon is known as the Fu-Kane formula~\cite{PhysRevB.76.045302}.  If time-reversal symmetric insulators are also endowed with the inversion symmetry, the product of parity eigenvalues at time-reversal invariant momenta (TRIMs) informs us of both the weak and strong $\mathbb{Z}_2$ indices. Similarly, in time-reversal breaking insulators with the $n$-fold rotation symmetry, the product of rotation eigenvalues determines the Chern number modulo $n$~\cite{PhysRevB.86.115112}.  Recently, this type of relations between the band topology and the symmetry representations in the band structure are further investigated and extended to wider class of spatial symmetries, i.e., 230 space groups~\cite{Po2017,Bradlyn:2017aa,1711.11050} and 1651 magnetic space groups~\cite{Watanabeeaat8685}, and to more general class of band topology including HOTIs~\cite{PhysRevLett.119.246402,Fang17,1711.11589}.  The generalized Fu-Kane formula, dubbed as ``symmetry indicator" of band topology, significantly reduces the effort of computing the topological indices and thus should be useful in the actual material search~\cite{Adrian1,Adrian2,1807.08756,1807.09744,1807.08756}. 

As we review in Sec.~\ref{sec2}, the symmetry indicator $X_{\textbf{BS}}$ for a given symmetry setting takes the form of a finite Abelian group $\prod_i\mathbb{Z}_{n_i}$ [see Eq.~\eqref{XBSform}].  The previous study~\cite{Po2017} computed this group structure and tabulated the results exhaustively for three AZ symmetry classes, A (no internal symmetries), AI (TRS with $\Theta^2=+1$ is added to A), and AII (TRS with $\Theta^2=-1$ is added to A), for every space group in $d$-dimensions ($d=1$, $2$, and $3$).  However, the physical meaning of each $\mathbb{Z}_n$ factor of $X_{\textbf{BS}}$ is not obvious from this table alone, and it has been the subject of recent follow-up studies. This problem has been addressed for class AI~\cite{1711.11050} and AII~\cite{1711.11049,1711.11589}, but has been left untouched for class A. The first major result of this work is to clarify the physical manifestation of the symmetry indicator in class A in detail, to the same level as done for class AI and AII.

Armed with the improved understanding on these three fundamental classes, we then proceed to the seven remaining AZ symmetry classes, motivated by the future application to identifying new topological superconductors. We first study the group structure of the symmetry indicator, and then clarify their physical meaning. There are several previous works along this line in some limiting cases~\cite{PhysRevLett.105.097001,PhysRevB.81.220504,1701.01944}, but our results are more comprehensive and unified in the sense it coherently covers all AZ symmetry classes altogether. We summarize our finings in the form of statements (i)-(viii) below. In particular, statement (ii) in Sec.~\ref{sec4} is the most fundamental result of this work, which says the group structure of symmetry indicators are not affected by the presence of PHS or CS, although their physical manifestation can be altered as clarified by other statements.

This paper is organized as follows.  In Sec.~\ref{sec2}, we review the general formulation and the definition of symmetry indicators. Section~\ref{sec3} clarifies the physical meaning of the symmetry indicators for class A.  In Sec.~\ref{sec4} we derive the symmetry indicators for other internal symmetry classes. The physical properties of each indicator for these symmetry classes are examined in Sec.~\ref{sec5}.  

\section{Review of symmetry indicators}
\label{sec2}
We start with reviewing the basics of the symmetry indicators.  Let us imagine a set of connected valence bands that is isolated from conduction bands by a nonzero band gap at least at high-symmetry momenta.   The band gap may vanish at generic momenta where the little group $G_{\vec{k}}$ is trivial.  Let $n_{\vec{k}}^\alpha$ be the number of irreducible representations $u_{\vec{k}}^\alpha$ appearing in the set of bands at each high-symmetry momentum $\vec{k}$.  Here, $\alpha$ labels irreducible representations of the little $G_{\vec{k}}$. One has to use projective representations (or, double-valued representations~\cite{Bradley}) for spinful electrons~\footnote{Sometimes the spinful case and the spinless case are not carefully distinguished in class A, but there are many differences in the structure of $\{\textbf{BS}\}$ and the indicator $X_{\textbf{BS}}$ in general, associated with the projective nature of the representations $u_{\vec{k}}^\alpha$. }.  Thanks to the assumed band gap at high-symmetry momenta, $n_{\vec{k}}^\alpha$ is well-defined even when the band gap closes at some generic momenta. Collecting $n_{\vec{k}}^\alpha$ for all inequivalent high-symmetry momenta and all irreducible representations, we can extract the set of integers $\vec{b}=\{n_{\vec{k}}^\alpha\}$, which contains useful information on the topology of the valence bands.

\begin{table}[t]
\caption{$X_{\textbf{BS}}$ for inversion symmetry in dimensions 1, 2 and 3.
\label{tab:inversion}}
\begin{tabular}{c|ccc}
\hline\hline
 & \,\,\,\,\,\,\,\,\,$\text{{\calligra p}}\bar{1}$ (1D)\,\,\,\,\,\,\,\,\, & \,\,\,\,\,\,\,\,\, $p\bar{1}$ (2D)\,\,\,\,\,\,\,\,\, & \,\,\,\,\,\,\,\,\,$P\bar{1}$ (3D)\,\,\,\,\,\,\,\,\, \\\hline
$X_{\textbf{BS}}$ & 1 & $\mathbb{Z}_2$ & $(\mathbb{Z}_2)^3\times\mathbb{Z}_4$\\
\hline\hline
\end{tabular}
\end{table}

The integers $\{n_{\vec{k}}^\alpha\}$ must satisfy several kinds of nontrivial constraints, the so-called compatibility relations.  Some of these constraints are intrinsic to the assumed space group and some of them originate purely from TRS in the case of class AI and AII.  See Refs.~\cite{Po2017,Watanabeeaat8685} for more details. Conversely, if all the compatibility relations are fulfilled, there exists a band structure with this combination of $\{n_{\vec{k}}^\alpha\}$. We write the set of all valid combinations as $\{\textbf{BS}\}$.

To diagnose the band topology based on the symmetry data $\{n_{\vec{k}}^\alpha\}$, we compare $\{n_{\vec{k}}^\alpha\}$ of our interest against those corresponding to atomic insulators~\cite{Po2017}.  In an atomic insulator, every electron is localized to a local atomic orbital without any hopping, forming a product state in the real space. Atomic insulators thus provide trivial combinations of $n_{\vec{k}}^\alpha$.  Listing up all possible atomic insulators by changing the type of filled atomic orbitals and their positions in each unit cell, one can find all trivial combinations of representations $\vec{a}=\{n_{\vec{k}}^\alpha\}$, which we denote by $\{\textbf{AI}\}$. If the band structure has a combination $\vec{b}=\{n_{\vec{k}}^\alpha\}\in\{\textbf{BS}\}$ that does not belong to $\{\textbf{AI}\}$, it must exhibit some sort of nontrivial topology since the mismatch of representations serves as an obstacle to adiabatically connect to any atomic insulator.

Finally, one can obtain a classification of nontrivial combinations of representations $\{n_{\vec{k}}^\alpha\}$ akin to $K$-theory by noting that both $\{\textbf{BS}\}$ and $\{\textbf{AI}\}$, when negative integers are allowed, form abelian groups of the same rank $d$.  Since atomic insulators are a special kind of band insulators~\cite{Po2017}, $\{\textbf{AI}\}$ is a subgroup of $\{\textbf{BS}\}$ so that the quotient group 
\begin{equation}
\label{XBSform}
X_{\textbf{BS}}=\{\textbf{BS}\}/\{\textbf{AI}\}=\mathbb{Z}_{n_1}\times \mathbb{Z}_{n_2}\times\cdots
\end{equation}
is well defined. This is what we call the symmetry indicator of band topology~\cite{Po2017}. 

\section{Interpretation of symmetry indicators for class A}
\label{sec3}
As explained in the introduction, the physical implication of symmetry indicators has been left unclear for class A (the class without any internal symmetries; hence no TRS is assumed), although this is the most elementary symmetry class among the 10 AZ  classes.  We address this issue in this section.

\subsection{Inversion symmetry}
\label{sec:inv}
Let us first develop some intuition using examples of spinful electrons symmetric under the $d$-dimensional lattice translation and the spatial inversion (Table~\ref{tab:inversion}).   We begin with one dimension and then proceed to higher dimensions.  To simplify the notation we will always set the lattice constant to be unity.

\subsubsection{1D}
The rod group $\text{{\calligra p}}\bar{1}$ is generated by the 1D translation and the spatial inversion $I$ with $I^2=1$.  Integers $n_{k_x}^\alpha$ ($\alpha=\pm$) count the number of even and odd parity eigenvalues at the two 1D TRIMs $k_x=0,\pi$.  The only compatibility condition in this case is $\sum_{\alpha=\pm}n_{0}^\alpha=\sum_{\alpha=\pm}n_{\pi}^\alpha$ that equals the total number of filled bands. It can be readily shown that $\{\textbf{BS}\}=\{\textbf{AI}\}$ $(\simeq\mathbb{Z}^3)$ so that the quotient $X_{\textbf{BS}}$ is trivial. The product of the inversion parities at TRIMs is related to the Berry phase $B$ quantized to $0$ or $\pi$: $\prod_{k_x=0,\pi}(-1)^{n_{k_x}^{-}}=e^{i B}~$\cite{PhysRevB.83.245132}. However, the fact that $X_{\textbf{BS}}=1$ implies that there is always a corresponding atomic limit regardless of $B=0$ or $\pi$. This conclusion is common among classes A, AI, and AII.

\subsubsection{2D}
The layer group $p\bar{1}$ is generated by the 2D translation together with the inversion $I$. The product of inversion parities at the four 2D TRIMs $\eta\equiv\prod_{k_x, k_y=0,\pi}(-1)^{n_{(k_x,k_y)}^{-}}$ distinguishes elements in $\{\textbf{AI}\}$ and $\{\textbf{BS}\}$: $\eta$ can be $\pm1$ in general but is constrained to be $+1$ for those corresponding to an atomic insulator. In fact, $\eta$ determines the Chern number modulo two, i.e., $\eta=(-1)^C$~\cite{PhysRevB.86.115112}, which explains the indicator $X_{\textbf{BS}}=\mathbb{Z}_2$ for this group. Note that nonzero Chern numbers are allowed only in the absence of TRS.  Therefore, $\eta=-1$ cannot be realized by insulators in class AI or AII --- $\eta=-1$ implies a Dirac semimetal in class AI~\cite{1711.11050} and $\eta$ is fixed to $+1$ due to the Kramers degeneracy in class AII. This difference in 2D results with/without TRS persists to 3D as we will see now.

\begin{figure}[t]
	\begin{center}
		\includegraphics[width=0.8\columnwidth]{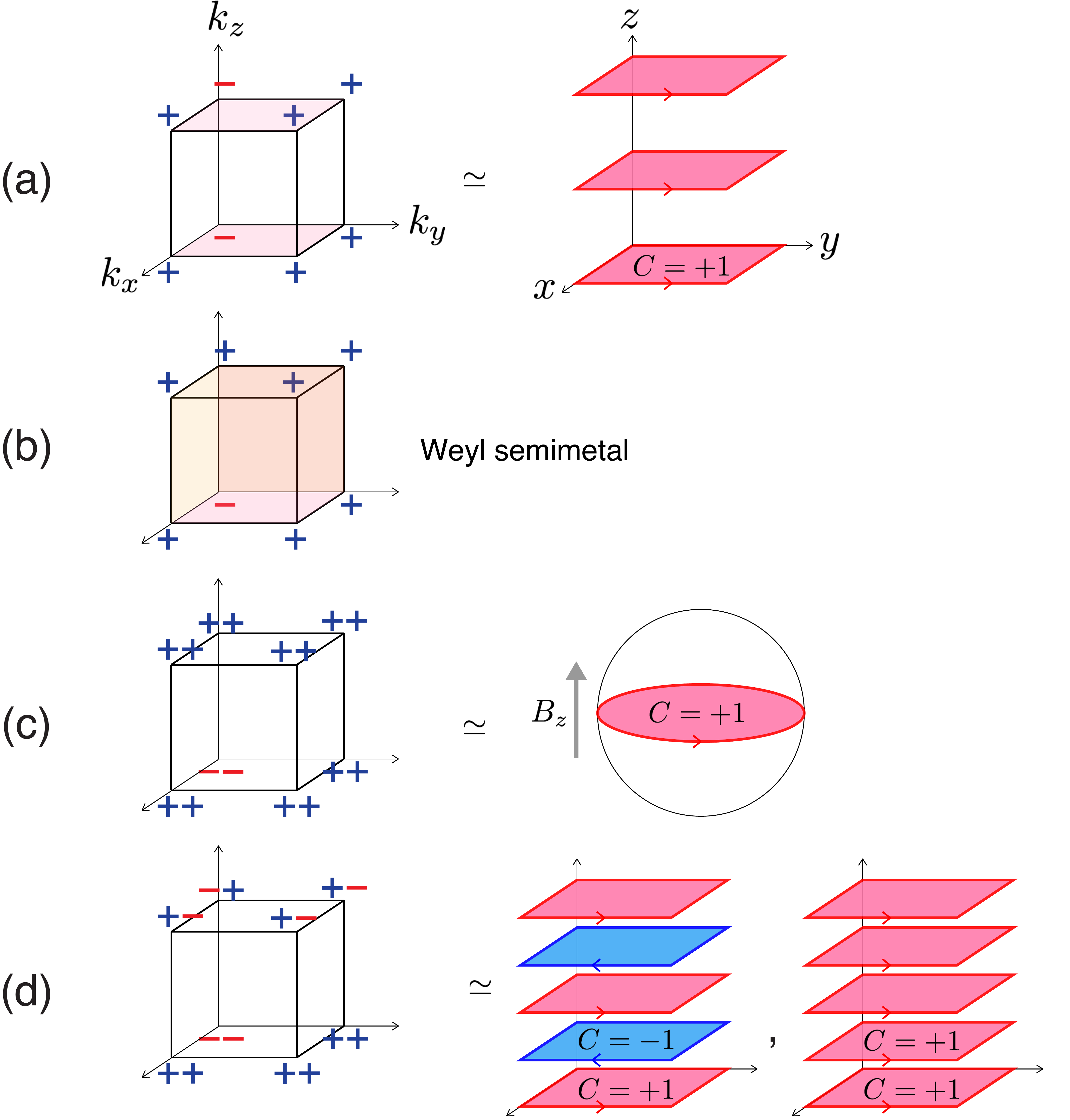}
		\caption{\textbf{Symmetry indicators for class A in $P\bar{1}$.} Planes with an odd Chern number are colored, while those with an even Chern number are not. (a) An example of parity eigenvalues in $(0,0,1;2)$ phase, which can be realized by stacking Chern insulators. (b) An example of $(0,0,0;3)$ phase corresponding to a Weyl semimetal.  (c) An example of $(0,0,0;2)$ phase obtained by applying an external magnetic field to a strong topological insulator. (d) Another example of $(0,0,0;2)$ phase, which can be realized as an antiferromagnetic Chern insulator and a ferromagnetic Chern insulator.
		\label{fig3Dinversion}}
	\end{center}
\end{figure}

\subsubsection{3D}
\label{sec:inv3D}
Finally, let us discuss the symmetry indicator for class A under the space group $P\bar{1}$ generated by 3D translation and the inversion $I$. The indicator is found to be $X_{\textbf{BS}}=(\mathbb{Z}_2)^3\times\mathbb{Z}_4$. The three factors of $\mathbb{Z}_2$ can be understood as the weak phases of stacked 2D Chern insulators discussed in the previous section [Fig.~\ref{fig3Dinversion} (a)], while the $\mathbb{Z}_4$ factor includes intrinsically 3D strong phases [Fig.~\ref{fig3Dinversion} (b)--(d)].  We can express elements of $X_{\textbf{BS}}$ as $(\nu_x,\nu_y,\nu_z;\mu_1)$ [$\nu_a=0, 1$ and $\mu_1=0,\ldots,3$].  Here, $(-1)^{\nu_a}$ ($a=x,y,z$) agrees with the product of the inversion parities at four TRIMs on the $k_a=\pi$ plane, and  $\mu_1$ is defined as a half of the \emph{sum}, not the product, of the inversion parities over the eight TRIMs: 
\begin{equation}
\mu_1\equiv\frac{1}{2}\sum_{\vec{k}\in\text{TRIMs}}(n_{\vec{k}}^+-n_{\vec{k}}^-)\in\mathbb{Z}. 
\end{equation}
In phases with an odd $\mu_1$ [Fig.~\ref{fig3Dinversion} (b)], the parity of the Chern number on the $k_z=0$ plane and that on the $k_z=\pi$ plane must be opposite. This forces the band gap to vanish somewhere in the Brillouin zone, realizing a Weyl semimetal~\cite{PhysRevB.85.165120}.   Again, this phase requires the absence of TRS.

The most nontrivial indicator is $(0,0,0;2)$, in which the band gap can be nonzero everywhere and the Chern number may vanish in all the fixed $k_a$ planes in the 3D Brillouin zone.  Even if this is the case, the insulator should still realize at least a HOTI exhibiting one-dimensional chiral hinge modes on the surface. The easiest way to see this is to perturb a strong topological insulator. In the presence of TRS, the parity combination shown in Fig.~\ref{fig3Dinversion} (c) implies the strong $\mathbb{Z}_2$ index~\cite{PhysRevB.76.045302}. If the TRS is weakly broken by an external magnetic field respecting the inversion symmetry and preserving the bulk gap, surface Dirac modes will be gapped out but there must be a one-dimensional domain wall at which the normal component of the magnetic field vanishes, hosting a chiral mode as illustrated in Fig.~\ref{fig3Dinversion} (c). A tight-binding model for this phase is presented in Eq.~\eqref{eq:HOTI} below~\cite{Matsugatani}.  Another way to see this is via an alternating stacking of Chern insulators with the $+1$ and $-1$ Chern numbers [see Fig.~\ref{fig3Dinversion} (d)], forming an antiferromagnetic Chern insulator~\cite{SitteRoschAltmanFritz, ZhangKaneMele,MongEssinMoore}.  When the translation symmetry along the $z$ axis is relaxed, pairs of anti-propagating chiral modes can gap out each other, leaving behind an odd number of chiral modes protected by the inversion symmetry.  Alternatively, one can explicitly construct a surface Dirac theory to see the behavior of the mass term under the inversion symmetry~\cite{Fang17,1711.11589}.  The symmetry indicators have ambiguities as stressed in Refs.~\cite{1711.11049,1711.11589} and the $(0,0,0;2)$ phase can also be a ferromagnetic stacking of Chern insulators [see Fig.~\ref{fig3Dinversion} (d)]. 

\subsection{Other space groups}
\label{sec:other}

To achieve the understanding of $X_{\textbf{BS}}$ for class A in other space groups in 3D, it is sufficient to discuss the seven \emph{key} space groups summarized in Table~\ref{tab:KeySG}, since those space groups not on this list (but with nontrivial $X_{\textbf{BS}}$) are either a supergroup of at least one of key space groups or some form of their variations (with a few minor exceptions discussed in Sec.~\ref{sec:Asum}). See Tables~\ref{tab:app1} and \ref{tab:app2} in the Appendix for the detailed grouping of 230 space groups.  Note that the choice of key space groups for class A differs from the choice in Ref.~\cite{1711.11049} for class AII, because the presence or absence of TRS fundamentally affects the possible topology and their relation to representations as we have seen through examples in Sec.~\ref{sec:inv}.  As we have already discussed $P\bar{1}$ in the previous section, we will discuss the rest of six key space groups one by one in the following.  Readers not interested in the details can skip to Sec.~\ref{sec:Asum}.

\begin{table}
\begin{center}
\caption{
\label{tab:KeySG} \textbf{Seven key space groups and their symmetry indicators.} The indicator $X_{\textbf{BS}}$ is split into the product of the weak indices $X_{\textbf{BS}}^{(w)}$ and the strong index $X_{\textbf{BS}}^{(s)}$.  The superscript ``Ch" means a stacked Chern insulator and ``mCh" is a stacked mirror Chern insulator.
}
\begin{tabular}{c|c|c|c}\hline \hline
Key space group	& $X_{\textbf{BS}}^{(w)}$		& $X_{\textbf{BS}}^{(s)}$ 	& Strong Index		\\\hline
$P\bar{1}$	& $[\mathbb{Z}_2^{(\text{Ch})}]^3$ &$\mathbb{Z}_4$ & $\mu_1$\\
$Pn$	 		& $\mathbb{Z}_n^{(\text{Ch})}$	 &		1		& 	-- \\
$Pn/m$		& $[\mathbb{Z}_n^{(\text{mCh})}]^2$& $\mathbb{Z}_n$		& mCh\\
$P\bar{4}$	& $\mathbb{Z}_4^{(\text{mCh})}$& $\mathbb{Z}_2\times\mathbb{Z}_2$ & Ch, $\mu_4$\\
$Pmmm$ (spinful)		& $[\mathbb{Z}_2^{(\text{mCh})}]^3$ & $\mathbb{Z}_4$ &  $\kappa_1$\\
$P4/mmm$ (spinful)		& $\mathbb{Z}_2^{(\text{mCh})}\times\mathbb{Z}_4^{(\text{mCh})}$ & $\mathbb{Z}_8$	& $\Delta$	\\
$Pcc2$ (spinless)		&  1 &$\mathbb{Z}_2$					& $\mu_2$\\
\hline\hline
\end{tabular}
\end{center}
\end{table}

\subsubsection{$Pn$}
Space group $Pn$ ($n=2,3,4,6$) is generated by $n$-fold rotation about the $z$ axis and the translation symmetry. On each fixed $k_z$ plane, rotation eigenvalues diagnose the Chern number modulo $n$~\cite{PhysRevB.86.115112}. Correspondingly, $X_{\textbf{BS}}=\mathbb{Z}_n$.  The generator of this group can be given by stacking Chern insulators with $C=1$.

\subsubsection{$Pn/m$}
Space group $Pn/m$ ($P2/m$, $P\bar{6}$, $P4/m$, and $P6/m$) has an additional mirror plane $M_z$ orthogonal to the rotation axis of $Pn$.  In this case, one can define a mirror Chern number (mCh) $C_{\pm}^{k_z}$ on the $k_z=0$ and the $k_z=\pi$ plane for each mirror sector.  Two weak mirror Chern insulators (for the two mirror eigenvalues) account for two factors of $\mathbb{Z}_n$. The remaining $\mathbb{Z}_n$ factor is a strong phase that can be time-reversal symmetric, in which the mirror Chern number on the $k_z=0$ plane is $C_{+}^{0}=-C_{-}^{0}\neq0$ while those on the $k_z=0$ plane vanish. 

\subsubsection{$P\bar{4}$}
Space group $P\bar{4}$ is generated by the four-fold rotoinversion $S_4$ (the four-fold rotation followed by the inversion) and the translation.  The rotoinversion $S_4$ also diagnoses the Chern number mod 4 and it gives a weak Chern insulator, corresponding the $\mathbb{Z}_4$ factor in $X_{\textbf{BS}}$, in which the Chern numbers on the $k_z=0$ and the $k_z=\pi$ plane are the same.  Since $S_4$ is not a symmetry unless $k_z=0$ or $\pi$, the Chern number on these two planes can differ by $4n+2$. In this case there must be Weyl points somewhere in between $0<k_z<\pi$. This phase corresponds to one of the two $\mathbb{Z}_2$ factors in $X_{\textbf{BS}}$. 

The remaining $\mathbb{Z}_2$ factor is related to a HOTI. This phase can be characterized by the sum of the $S_4$-eigenvalues of valence bands at the four $S_4$ invariant momenta $K_4$: $(0,0,0)$, $(\pi,\pi,0)$, $(0,0,\pi)$, and $(\pi,\pi,\pi)$.  For example, in the case of spinful electrons, the possible values of $S_4$-eigenvalues are $e^{\frac{\alpha\pi}{4}i}$ ($\alpha=1,3,5,7$), and denote by $n_K^\alpha$ the number of valence bands with the eigenvalue $e^{\frac{\alpha\pi}{4}i}$ at momentum $\vec{k}\in K_4$. 
\begin{equation}
\mu_4\equiv\frac{1}{\sqrt{2}}\sum_{\vec{k}\in K_4}\sum_{\alpha}e^{\frac{\alpha\pi}{4}i}n_K^\alpha.
\end{equation}
Atomic insulators can change the value of $\mu_4$ only by $2\pm 2i$. Therefore, insulators with $\mu_4=2$ or $2i$, for example, must be nontrivial. We can construct an insulator with this indicator by an antiferromagnetic stacking of Chern insulators, just as we have discussed for $P\bar{1}$. When the translation symmetry is broken while keeping the $S_4$ symmetry, the insulator must exhibit an odd number of chiral 1D modes on the surface.

\subsubsection{$Pmmm$ for spinful electrons}
Space group $Pmmm$ for spinful electrons has the same form of indicators as $P\bar{1}$, i.e., $X_{\textbf{BS}}=(\mathbb{Z}_2)^3\times\mathbb{Z}_4$, but their interpretations are different. $Pmmm$ has three orthogonal mirror symmetries $M_x$, $M_y$, and $M_z$ in addition to $P\bar{1}$.  All the representations at TRIMs become two dimensional due to the algebra $M_xM_y=-M_yM_x=M_z$, effectively forming Kramers' pairs (i.e., each band is paired with another one with the same inversion parity but the opposite mirror eigenvalue at TRIMs). Therefore, one can use the index for class AII introduced in Refs.~\cite{1711.11049,1711.11589}:
\begin{equation}
\kappa_1\equiv\frac{1}{4}\sum_{\vec{k}\in\text{TRIMs}}(n_{\vec{k}}^+-n_{\vec{k}}^-)\in\mathbb{Z}.
\end{equation}
The generator of each $\mathbb{Z}_2$ factor is given by stacking mirror Chern insulators.  Insulators with an odd $\kappa_1$ can be seen as the $\mathbb{Z}_2$ strong topological insulator, despite the absence of TRS, since the mirror Chern number on the $k_z=0$ plane ($C_{+}^{0}=-C_{-}^{0}$) and that on the $k_z=\pi$ plane ($C_{+}^{\pi}=-C_{-}^{\pi}$) must be different.  The $(0,0,0;2)$ phase can be a HOTI with helical hinge modes~\cite{1711.11049,1711.11589}.

\subsubsection{$P4/mmm$ for spinful electrons}
Space group $P4/mmm$ has an indicator $X_{\textbf{BS}}=\mathbb{Z}_2\times\mathbb{Z}_4\times\mathbb{Z}_8$.  
The first two factors, $\mathbb{Z}_2$ and $\mathbb{Z}_4$, can be readily accounted by its subgroup $Pmmm$ and $P4/m$, respectively. The $\mathbb{Z}_2$ factor is the common mirror Chern number in the $xz$ plane and the $yz$ plane.  The generator of the $\mathbb{Z}_4$ has a mirror Chern number in $xy$ plane $C_{+}^{0}=C_{+}^{\pi}=-C_{-}^{0}=-C_{-}^{\pi}=1$ mod 4.

To characterize the remaining $\mathbb{Z}_8$ factor, note that $P4/mmm$ also contains both $P\bar{1}$ and $P\bar{4}$ as subgroups.  Because of the three orthogonal mirrors $M_x$, $M_y$ and $M_z$, all the representations effectively form Kramers' pairs. As we discussed in the previous section, in this case $\mu_1$ for $P\bar{1}$ is enhanced to $\kappa_1$. Similarly, $\mu_4$ is enhanced to $\kappa_4$, introduced for class AII system in Refs.~\cite{1711.11049,1711.11589}:
\begin{eqnarray}
\kappa_4\equiv\frac{1}{2\sqrt{2}}\sum_{K\in K_4}\sum_{\alpha}e^{\frac{\alpha\pi}{4}i}n_K^\alpha\in\mathbb{Z}.
\end{eqnarray}
The index for the $\mathbb{Z}_8$ factor is given by $\Delta\equiv\kappa_1-2\kappa_4$ mod 8.  $\Delta$ mod 4 agrees with the difference  of mirror Chern number on the $k_z=0$ plane and the $k_x=\pi$ plane. In insulators with $\Delta=4$ mod 8, the mirror Chern numbers on the $k_x=0,\pi$ planes can all vanish. Even in that case the insulator still exhibits a helical edge mode on its surface as discussed in Ref.~\cite{1711.11589}.

\subsubsection{$Pcc2$ for spinless electrons}
Space group $Pcc2$ is generated by two-fold rotation $C_{2z}: (x,y,z)\mapsto(-x,-y,z)$, the glide symmetry $G_{z}: (x,y,z)\mapsto(x,-y,z+\frac{1}{2})$, and the lattice translation. The indicator is $X_{\textbf{BS}}=\mathbb{Z}_2$ for spinless electrons, while it is trivial for spinful electrons.  To understand this, note that $G_z$ and $C_{2z}$ commute for the spinless case. As a result, irreducible representations are all one-dimensional and are given by $C_{2z}=\xi_1$ and $G_{z}=\xi_2 e^{-ik_z/2}$ ($\xi_1=\pm1$ and $\xi_2=\pm1$) on every high-symmetry line [$(0,0,k_z)$, $(\pi,0,k_z)$, $(0,\pi,k_z)$, and $(\pi,\pi,k_z)$]. Because of the nonsymmorphic nature of $G_z$, the representation $(\xi_1,\xi_2)$ has to be paired with $(\xi_1,-\xi_2)$, so that each rotation eigenvalue appears an even number of times.  Namely, if we denote by $n_{\vec{k}}^{\pm}$ the number of $\pm1$ eigenvalues of $C_{2z}$ on each high-symmetry line, $n_{\vec{k}}^{\pm}$'s are all even. Given this, we define
\begin{equation}
\mu_2\equiv\frac{1}{4}\sum_{\vec{k}: \text{TRIMs at $k_z=0$}}(n_{\vec{k}}^+-n_{\vec{k}}^-)\in\mathbb{Z}.
\end{equation}
This index is always even for atomic insulators. 
One way of realizing an insulator with an odd $\mu_1$ is to start with a Chern insulator with $C=+1$ on the $z=0$ plane.  The glide operation $(G_z)^n$ ($n\in\mathbb{Z}$) will convert this Chern insulator to a Chern insulator with $C=(-1)^n$ on the $z=n/2$ plane.  Gathering all of these planes, we get an antiferromagnetic stacking of Chern insulators with nontrivial $\mathbb{Z}_2$ index protected by glide symmetry~\cite{PhysRevB.91.155120}.

\subsection{Summary}
\label{sec:Asum}
After all, above discussions suggest that 
\begin{quote}
(i) \textit{In 2D, every $X_{\textbf{BS}}$ nontrivial phases in class A is either a Chern insulator or a mirror Chern insulator. The statement also applies to weak phases in 3D corresponding to stacked 2D insulators (see the column $X_{\textbf{BS}}^{(w)}$ in Table~\ref{tab:KeySG}).}
\end{quote}
In contrast, there are variety of intrinsically 3D phases indicated by $X_{\textbf{BS}}^{(s)}$ in Table~\ref{tab:KeySG}.
Some strong phases in 3D are Weyl semimetal (e.g., insulators with an odd $\mu_1$ in $P\bar{1}$ and for a $\mathbb{Z}_2$-indicator in $P\bar{4}$). Some others are strong mirror Chern insulators in which the mirror Chern numbers in the $k_z=0,\pi$ planes are different (e.g., insulators with an odd $\kappa_1$ in $Pmmm$ and for a $\mathbb{Z}_n$-insulator in $Pn/m$).  There are also HOTI phases with either chiral edge modes [$(0,0,0,2)$ phases for $P\bar{1}$ and $\mu_4$ nontrivial phases in $P\bar{4}$] or helical edge modes [$(0,0,0,2)$ phase for $Pmmm$ and $\Delta=4$ phase in $P4/mmm$].

For spinless electrons breaking the TRS in 3D, there are also a few cases not covered by the above discussions. In space groups $P4_2bc$ (No.~106) and $I4_1cd$ (No.~110), all atomic insulators have the filling $\nu=4n$ while nontrivial band insulators may have the filling $\nu=4n+2$. Hence, the filling alone plays the role of the $\mathbb{Z}_2$ indicator~\cite{SciAdv,Po2017}. The physical property of this phase (the so-called ``filling-enforced quantum band insulator") has not been understood yet.

\section{Symmetry indicators for other symmetry classes} 
\label{sec4}

\begin{figure}[t]
	\begin{center}
		\includegraphics[width=0.99\columnwidth]{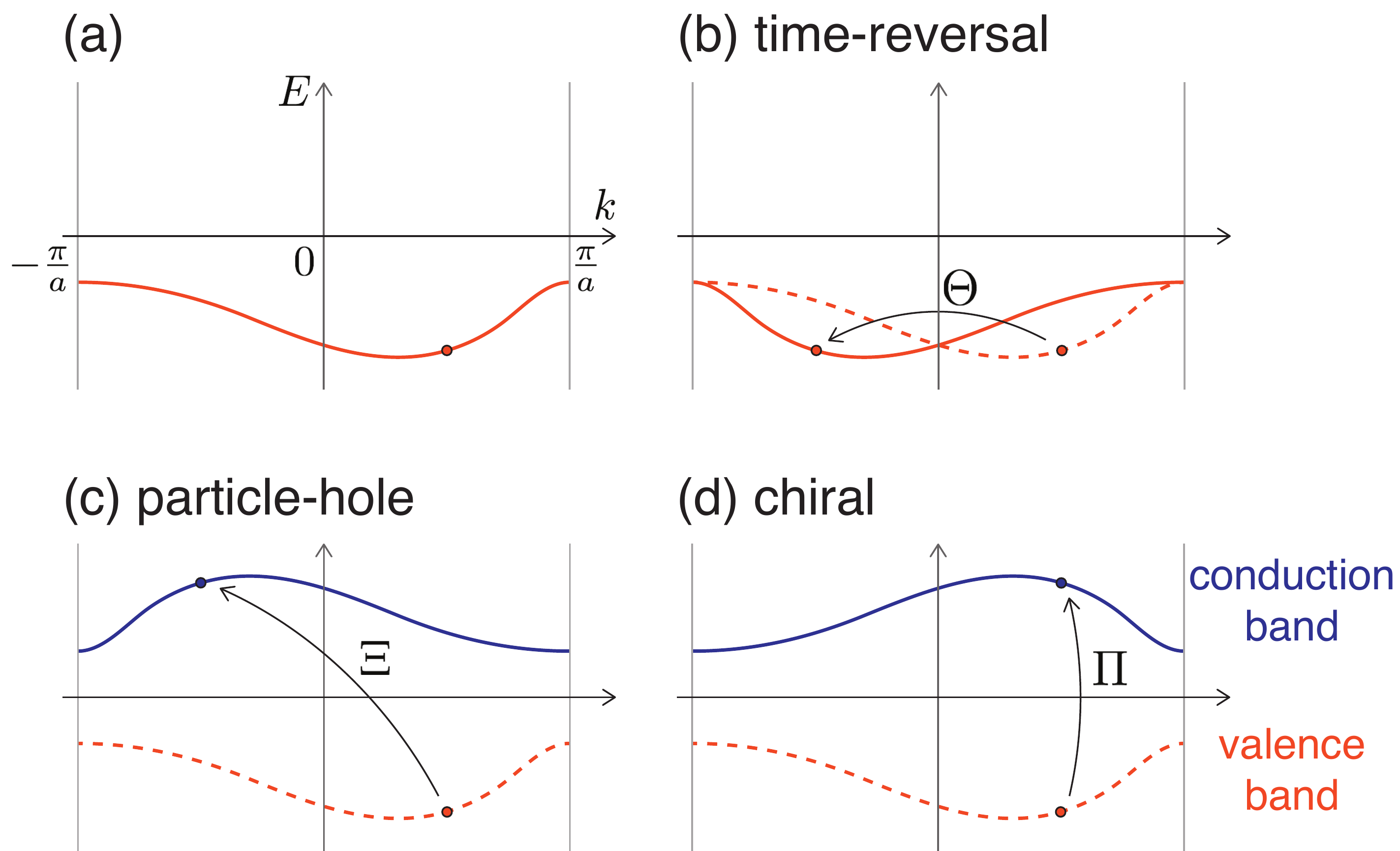}
		\caption{\textbf{The action of internal symmetries on a valence band.} (a) The original band structure. (b) The time-reversal symmetry $\Theta$. (c) The particle-hole symmetry $\Xi$. (d) The chiral symmetry $\Pi$.\label{figAction}}
	\end{center}
\end{figure}

Now that we have understood the physical meaning of the most of the indicators for class A as summarized, let us ask what happens if we add internal symmetries to the above discussion. We examine the effect of the TRS $\Theta$, the PHS $\Xi$, and the CS $\Pi$ one by one.  To simplify the discussion here we will always assume the chemical potential $\mu$ to be particle-hole symmetric, i.e., $\mu=0$.

Let us start with the TRS, which is antiunitary and satisfies $\Theta H_{\vec{k}}=H_{-\vec{k}}\Theta$. If $|u_{\vec{k}}\rangle$ denotes the Bloch state with the momentum $\vec{k}$ and the energy $\epsilon_{\vec{k}}<0$, the state $\Theta|u_{\vec{k}}\rangle$ has the inverted momentum $-\vec{k}$ but the same energy $\epsilon_{\vec{k}}<0$ [Fig.~\ref{figAction} (b)]. Depending on the value of $\Theta^2=\pm1$, $|u_{\vec{k}}\rangle$ and $\Theta|u_{\vec{k}}\rangle$ may or may not belong to the same band, but in either case $\Theta$ maps a valence band to a valence band.  As a result, adding $\Theta$ may increase the compatibility relations imposed on the combinations of representations $\{n_{\vec{k}}^\alpha\}$. Therefore, $\{\textbf{BS}\}$, $\{\textbf{AI}\}$, and $X_{\textbf{BS}}$ can be all altered in the presence of $\Theta$, and one has to separately work out for class A, AI, and AII one by one.  
This calculation is performed exhaustively for all 230 space groups in Ref.~\cite{Po2017} and the results are indeed different among the three symmetry settings.

\begin{table}[b]
\caption{Grouping the 10 Altland-Zirnbauer symmetry classes based on their time-reversal properties.  $K$ represents the complex conjugation.
\label{tab:AZ}}
\begin{tabular}{c|ccc}
\hline\hline
Additional symm. & $\,\,\,\,$ A (no $\Theta$) $\,\,\,\,$	& $\,\,$ AI $(\Theta=K)$ $\,\,$	& AII	($\Theta=i\sigma_2K$)	 \\\hline
$\Xi=\tau_1 K$ &D		& BDI		& DIII	\\
$\Xi=i\tau_2 K$ &C		& CI			& CII		\\
$\Pi=\tau_1$ & AIII		& --			& --	 	\\
\hline\hline
\end{tabular}
\end{table}

The remaining question is whether one has to repeat the same calculation for other seven symmetry classes in Table~\ref{tab:AZ}. Here we argue that one does not have to, since none of $\{\textbf{BS}\}$, $\{\textbf{AI}\}$, or $X_{\textbf{BS}}$ is affected by adding either the PHS $\Xi$ (antiunitary) or the CS $\Pi$ (unitary).  To see this, note that $\Xi$ and $\Pi$, individually, interchange a valence band and a conduction band.  As suggested by $\Xi H_{\vec{k}}=-H_{-\vec{k}}\Xi$ and $\Pi H_{\vec{k}}=- H_{\vec{k}}\Pi$, the Bloch state $\Xi|u_{\vec{k}}\rangle$ has the momentum $-\vec{k}$ and the energy $-\epsilon_{\vec{k}}>0$ [Fig.~\ref{figAction} (c)], while $\Pi|u_{\vec{k}}\rangle$ has the momentum $\vec{k}$ and the energy $-\epsilon_{\vec{k}}>0$ [Fig.~\ref{figAction} (d)].  Therefore, neither $\Xi$ or $\Pi$ enhances the constraints on the combinations of representations $\{n_{\vec{k}}^\alpha\}$ of \emph{valence} bands.  Thus, 
\begin{quote}
(ii) \emph{Symmetry classes with the same TR property (i.e., in the same same column in Table~\ref{tab:AZ}) have the identical indicators.}
\end{quote}
Combining (ii) with the fact that $X_{\textbf{BS}}$ for class A, AI, and AII in 1D are all trivial in 1D~\cite{Po2017}, we conclude
\begin{quote}
(iii) \emph{There does not exist any symmetry indicator in 1D in any symmetry class.}
\end{quote}
In other words, the combination of representations alone can never diagnose the nontrivial entries of the topological periodic table in 1D, such as the $\mathbb{Z}_2$ index for class D.

Note that the way we formulated the problem always starts with the Hamiltonian $H_{\vec{k}}$ in the class A, AI, or AII and then adds either the CS or the PHS symmetry. Even the BCS Hamiltonian for superconductors that explicitly breaks the electron number conservation possesses a U(1) symmetry when written in the Nambu representation. As far as the symmetry representations are concerned this formulation is the most convenient.

\section{Interpretation of symmetry indicators}
\label{sec5}
Here we discuss the properties of the phases indicated by $X_{\textbf{BS}}$ in the presence of the additional CS or PHS.

\subsection{Class AIII in 2D}
We have shown in the previous section that the symmetry indicators $X_{\textbf{BS}}$ for classes A and AIII, for instance, are the same. This immediately raises the following question. As we have seen above, almost all indicators for class A are associated with Chern numbers.  However, nonzero Chern numbers are prohibited in class AIII as suggested by the absence of a $\mathbb{Z}$ factor in the topological periodic table for class AIII in 2D. This statement can also be readily derived by combining the two facts: (i) the sum of the Chern numbers for the valence bands and the conduction bands must vanish ($C_v+C_c=0$) and (ii) the Chern number for the valence bands and the conduction bands must be the same ($C_v=C_c$) due to the CS $\Pi$. If indicators are the same despite this apparent difference between class A and AIII, what kind of topology does the nontrivial indicator imply in the presence of the CS?

To answer this question, recall that we did not assume a band gap at generic momenta at which the little group $G_{\vec{k}}$ is no larger than just the translation subgroup of the space group.  Suppose that the combination of representations $\{n_{\vec{k}}^\alpha\}$ is such that it implies a nonzero Chern number as far as the band gap does not vanish over the entire Brillouin zone.  Then the only way the valence band can evade the prohibited Chern number is via gap closing somewhere inside of the Brillouin zone.  Combining this argument with (i) and (ii), we find that
\begin{quote}
(iv) \emph{In 2D, Every $X_{\textbf{BS}}$ nontrivial band structure for class AIII must be gapless. The statement also applies to weak phases in 3D.}
\end{quote}

\begin{figure}[t]
	\begin{center}
		\includegraphics[width=0.99\columnwidth]{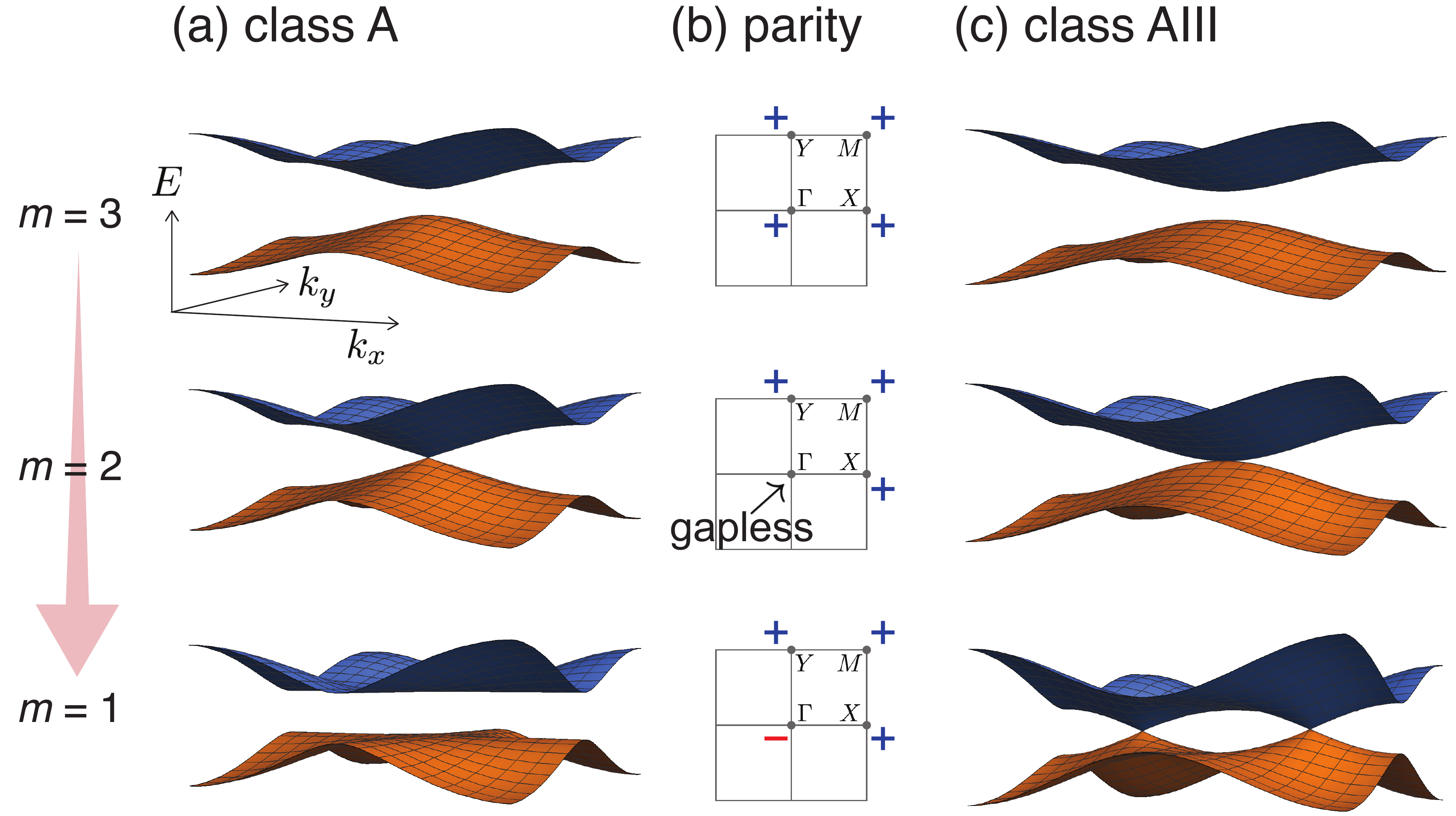}
		\caption{\textbf{Comparison of class A and AIII.} (a) The band dispersion of the model in Eq.~\eqref{TBmodelA} for $m=3$, $2$, and $1$. (b) The corresponding parity eigenvalues of the lower band. (c) The same as (a) but for the model in Eq.~\eqref{TBmodelAII}.\label{figAIII}}
	\end{center}
\end{figure}

To demonstrate our argument through a simple exercise, let us start with a model symmetric under the layer group $p\bar{1}$:
\begin{equation}
H_{\vec{k}}^{\text{A}}=-\sin k_x \tau_x-\sin k_y \tau_y-(m-\cos k_x-\cos k_y) \tau_z,\label{TBmodelA}
\end{equation}
in which the inversion symmetry $I$ is represented by $\tau_z$. Here, $\tau_a$ ($a=x,y,z$) are Pauli matrices (we reserve $\sigma_i$ for Pauli matrices associated with spin). 
As explained above, the product of the parity eigenvalues over the four TRIMs $\Gamma=(0,0)$, $X=(\pi,0)$, $Y=(0,\pi)$, $M=(\pi,\pi)$ determines the Chern number modulo two.

For the model in Eq.~\eqref{TBmodelA}, the parity eigenvalues at all TRIMs are $+1$ when $m>2$. As $m$ decreases, a band inversion occurs when $m=2$ and the parity eigenvalue at $\Gamma$ becomes $-1$ while those at other TRIMs stay unchanged for $0<m<2$. Consequently, the Chern number of the lower band jumps from $0$ to $1$ in this process.  The band dispersion for $m=3$, $2$, and $1$ are illustrated in Fig.~\ref{figAIII} (a), and the corresponding inversion parities are shown in Fig.~\ref{figAIII} (b).

Now let us add the CS $\Pi=\tau_x$ to the model by dropping the term proportional to $\tau_x$:
\begin{equation}
H_{\vec{k}}^{\text{AIII}}=-\sin k_y \tau_y-(m-\cos k_x-\cos k_y) \tau_z,\label{TBmodelAII}
\end{equation}
The inversion parities at TRIMs are unchanged from the previous model and the product of parity eigenvalues becomes $-1$ when $|m|\leq1$. In this range of the parameter the band structure becomes gapless as shown in Fig.~\ref{figAIII} (c). This is precisely what we predicted above --- the would-be Chern insulator for class A becomes gapless due to the added CS that prohibits a nonzero Chern number. This line of reasoning was recently used to prove that all the $X_{\textbf{BS}}$ nontrivial phases have nodal points or nodal lines in class AI~\cite{1711.11050}.

\subsection{Class AIII in 3D}
As a more nontrivial example, let us consider the following 3D models:
\begin{eqnarray}
\label{eq:HOTI}
H_{\vec{k}}^{\text{A}}&=&-\sin k_x \tau_x\sigma_x-\sin k_y \tau_x\sigma_y-\sin k_z \tau_x\sigma_z\\
&&-(2-\cos k_x-\cos k_y-\cos k_z)\tau_z\sigma_0-B_z \tau_0\sigma_z.\notag
\end{eqnarray}
This model is precisely the inversion symmetric topological insulator under a uniform magnetic field $B_z$ discussed in Sec.~\ref{sec:inv3D}.   The inversion symmetry is represented by $I=\tau_z\sigma_0$ and the TRS $\mathcal{T}=-i\tau_0\sigma_y K$ is broken by the magnetic field.  As far as $|B_z|<1$, the parity eigenvalues of valence bands are identical to those in Fig.~\ref{fig3Dinversion} (c) and the model realizes a HOTI with a chiral edge mode, in which the 3D bulk and 2D surfaces are completely gapped, as illustrated in Fig.~\ref{fig3Dinversion} (c). 

\begin{figure}[t]
	\begin{center}
		\includegraphics[width=1.0\columnwidth]{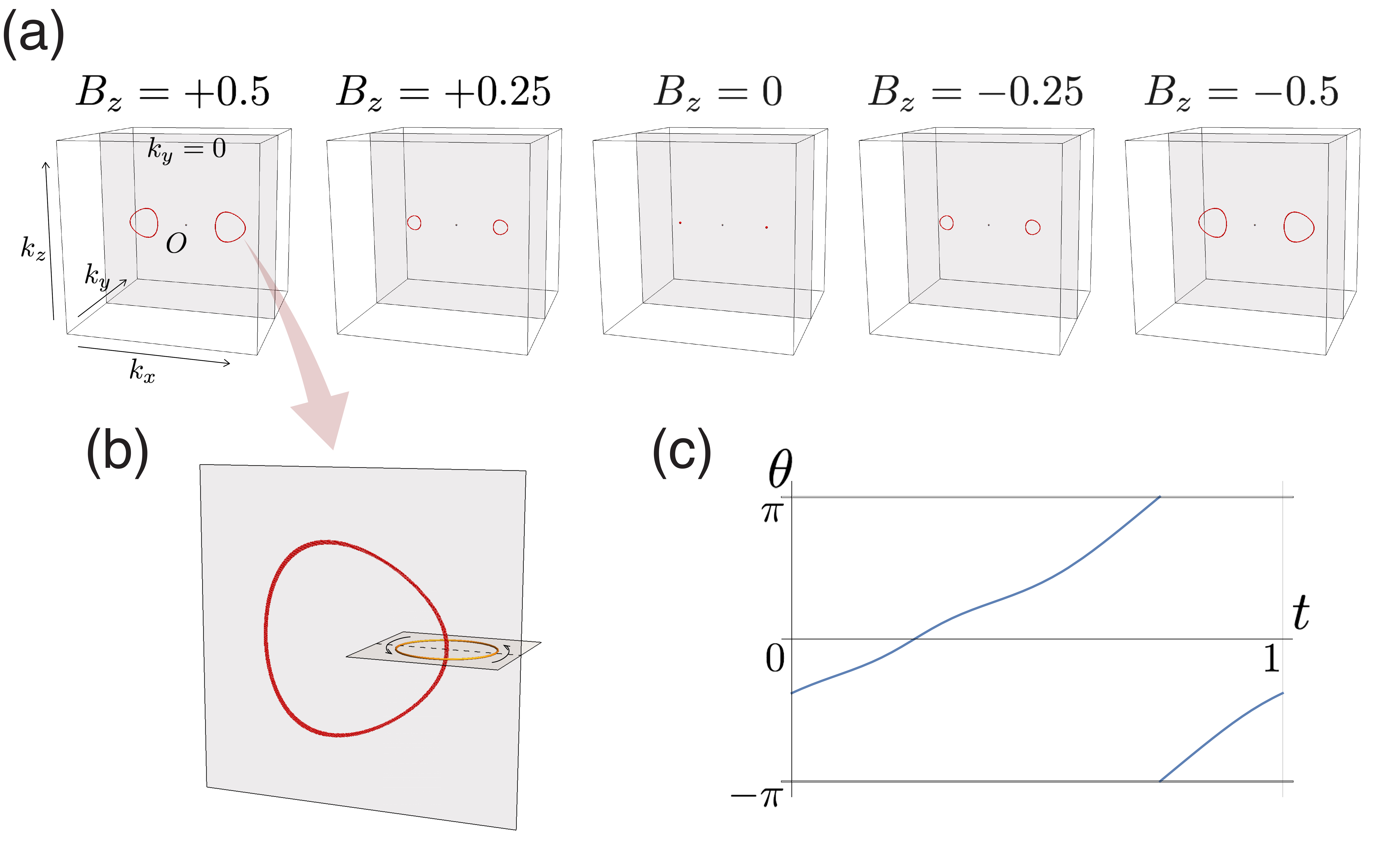}
		\caption{\textbf{Nodal line semimetals in class AIII.}  (a) The nodal lines in the model Eq.~\eqref{eq:HOTI2} are plotted for different values of $B_z$.  (b) A zoom-up of a nodal ring for $B_z=0.5$. The yellow circle is the path for computing the winding. (c) The nontrivial winding of $\theta(t)$ [see Eq.~\eqref{theta}] associated with the nodal ring. \label{fig:nodes}}
	\end{center}
\end{figure}

Now let us introduce the CS to $H_{\vec{k}}^{\text{A}}$. First we use $\Pi=\tau_x\sigma_x$. Again dropping the first term in $H_{\vec{k}}^{\text{A}}$, we get
\begin{eqnarray}
\label{eq:HOTI2}
H_{\vec{k}}^{\text{AIII}}&=&-\sin k_y \tau_x\sigma_y-\sin k_z \tau_x\sigma_z\\
&&-(2-\cos k_x-\cos k_y-\cos k_z)\tau_z\sigma_0-B_z \tau_0\sigma_z.\notag
\end{eqnarray}
The would-be HOTI, exhibiting a chiral edge mode, can be smoothly deformed to a 2D Chern insulator~\cite{Matsugatani}, but such a phase is strictly prohibited in class AIII. Therefore the bulk band gap has to vanish somewhere in the Brillouin zone, forming nodal lines protected by the CS (just in the same way as the product of the inversion symmetry and the TRS for spinless electrons does).  Indeed, the band structure of the current model hosts two nodal rings as shown in Fig.~\ref{fig:nodes} (a). Intriguingly, the two rings are not removable even when they are shrunk to a point, as required by the above argument.

To see the stability of nodal rings more explicitly, let us demonstrate the nontrivial topology inherited to them, following Ref.~\onlinecite{PhysRevB.96.155105}.
To this end, we choose a new basis in which the CS and the Hamiltonian take the form
\begin{eqnarray}
\Pi=\begin{pmatrix}\openone&0\\0&-\openone\end{pmatrix},\quad H_{\vec{k}}^{\text{AIII}}=\begin{pmatrix}0&q_{\vec{k}}\\q_{\vec{k}}^\dagger&0\end{pmatrix}.
\end{eqnarray}
We take a circle around one of the two nodal rings, illustrated in Fig.~\ref{fig:nodes} (b), and parametrize as $\vec{k}(t)$ with $t\in[0,1]$. Along this circle, $\det q_{\vec{k}(t)}$ does not vanish and 
\begin{equation}
\label{theta}
\theta(t)=\text{Im}\log \det q_{\vec{k}(t)}
\end{equation}
is well-defined (modulo $2\pi$). The nodal ring is topologically protected by the nontrivial winding of $\theta(t)$ as a function of $t\in[0,1]$~\cite{PhysRevB.96.155105} as demonstrated in Fig.~\ref{fig:nodes} (c).

Now let us use another representation $\Pi=\tau_y\sigma_0$ of the CS. To respect this CS, we have to drop the magnetic field term and get
\begin{eqnarray}
\label{eq:HOTI3}
H_{\vec{k}}^{\text{AIII}}{}'&=&-\sin k_x \tau_x\sigma_x-\sin k_y \tau_x\sigma_y-\sin k_z \tau_x\sigma_z\\
&&-(2-\cos k_x-\cos k_y-\cos k_z)\tau_z\sigma_0.\notag
\end{eqnarray}
This time the surface Dirac mode cannot be gapped out by the magnetic field, since the term $B_z \tau_0\sigma_z$ is not compatible with the current choice of the CS.  Therefore, we cannot identify the insulating phase as a HOTI, and the bulk dispersion can remain gapped unlike the previous case.

\subsection{Other symmetry classes}
Let us discuss the remaining symmetry classes. First of all, since $X_{\textbf{BS}}$ for class BDI and CI are the same as class AI and since all phases with nontrivial indicators are gapless in class AI~\cite{1711.11050}, it immediately follows that
\begin{quote}
(v) \emph{Every $X_{\textbf{BS}}$ nontrivial band structure for class BDI and CI must be gapless just as in class AI.}
\end{quote}
Our conclusion that the band structure cannot have a full band gap whenever the indicator is nontrivial in these symmetry classes also implies a no-go theorem:
\begin{quote}
(vi) \emph{There does not exist any symmetry indicator that diagnoses the gapped phases in AI, BDI, or CI. The same is true for class AIII in 2D.}
\end{quote}

We can readily generalize the discussion to other symmetry settings. For example, the Chern number is constrained to be even in class C. 
In class CII, 
the $\mathbb{Z}_2$ quantum spin Hall index must be trivial~\cite{PhysRevB.78.195125,doi:10.1063/1.3149495,ryu2010topological}. Therefore, 
\begin{quote}
(vii) \emph{Every band structure for class C and CII is gapless if its $X_{\textbf{BS}}$ indicates a prohibited topological index, i.e., any odd Chern number for class C and the nontrivial quantum spin Hall index for class CII.}
\end{quote}

Finally, class D is compatible with arbitrary integral Chern number, and class DIII supports the nontrivial $\mathbb{Z}_2$ quantum spin Hall phase.  Therefore,
\begin{quote}
(viii) \emph{The same symmetry indicator for class A (AII) diagnoses the topological phases in class D (DIII).}
\end{quote}
This observation is consistent with earlier works~\cite{PhysRevLett.105.097001,PhysRevB.81.220504,1701.01944}.

\section{Discussions}
\label{sec6}
In this work, we generalized the previous studies on symmetry indicators to the 10 AZ symmetry classes in each of 230 space groups all together. We derived several kinds of general statements that predict topological semimetals or topological superconductors based on the symmetry representations of the valence bands.  

An obvious future work will be covering all 1651 magnetic space groups.  For each magnetic space group and for each choice of whether electrons have spin (which determines $\Theta^2=\pm1$), there can still be four symmetry settings: (i) without PHS or CS, (ii) with PHS ($\Xi^2=+1$), (iii) with PHS ($\Xi^2=-1$), and (iv) with CS but no PHS.  Our argument that adding PHS or CS does not affect the symmetry indicators still holds, but the interpretation part requires more detailed analysis.

\begin{acknowledgements}
The authors would like to thank Y. Tada, H. Fujita, H. Katsura, R.-J. Slager, H. C. Po, E. Khalaf, Z. Song, and C. Fang for useful discussions. The authors are especially grateful to K. Shiozaki for a critical comment on the initial version of the draft.  
The work of H. W. is supported by JSPS KAKENHI Grant No. JP17K17678.
\end{acknowledgements}

\clearpage
\bibliography{references}

\appendix 

\section{Summary of symmetry indicator in class A}
As we discussed in Sec.~\ref{sec:other} in the main text, the symmetry indicators in each of the 230 space groups can be understood by using the knowledge on the seven key space groups. We summarize the results in Table~\ref{tab:app1} (for spinful electrons) and Table~\ref{tab:app2} (for spinless electrons). In these tables, we use the numbering schemes of space groups in the international tables~\cite{ITC}.

We explain how to read these tables using the example of space group No.~140 for spinful electrons. It is located in the middle of the fifth column in Table~\ref{tab:app1}. The third and the fourth column tell us that the indicator of this group is $X_{\textbf{BS}}=X_{\textbf{BS}}^{(w)}\times X_{\textbf{BS}}^{(s)}=1\times\mathbb{Z}_4$.  To see the meaning of this indicator, one should refer to the first and the second column.  The second column ($G_0$) says that this space group is a $t$-subgroup of $I4/m$ (No.~87) and the indicator can be fully understood in terms of the indicator for $I4/m$. The difference between $I4/m$ and $P4/m$ in the first column is only the translation symmetry --- the former has the additional body-centered translation symmetry which is not present in the latter.  The indicators for $I4/m$ can be understood by simply forgetting about this body-centered translation symmetry using the same indices as $P4/m$. After all, the $\mathbb{Z}_4$ factor of $X_{\textbf{BS}}$ for space group No.~140 is the strong mirror Chern insulator.

\begin{table*}
\begin{center}
\caption{\label{tab:app1}\textbf{Symmetry indicators for spinful electrons in class A for 230 space groups.} }
\begin{tabular}{c|c|cc|cc|c}\hline \hline
Key SG &  $G_0$ & weak & $X_{\textbf{BS}}^{(w)}$ & strong & $X_{\textbf{BS}}^{(s)}$ & $t$-supergroups of $G_0$ \\\hline
& & & $(\mathbb{Z}_2)^3$ & & $\mathbb{Z}_4$& 2\\
& &  & $\mathbb{Z}_2$ & & $\mathbb{Z}_4$& 148\\
\multirow{2}{*}{$P\bar{1}$}&\multirow{2}{*}{$P\bar{1}$} & \multirow{2}{*}{Ch} & $\mathbb{Z}_2$ & \multirow{2}{*}{$\mu_1$} & $\mathbb{Z}_2$& 12, 13, 15\\
& &  &   &  &  & 11, 14, 48--50, 52--54, 56--64, 66--68, 70, 72--74, 125,\\
& &  &1 &  & $\mathbb{Z}_2$ &  126, 129, 130, 133, 134, 137, 138, 141, 142, 162--167,\\ 
& &  & & & & 201, 203, 205, 206, 222, 224, 227, 228, 230\\\hline\hline
$P2$&$P2$ & Ch & $\mathbb{Z}_2$& -- &1 & 3, 171, 172\\\hline
&$P4$ & Ch & $\mathbb{Z}_4$& -- &1  & 75\\\cline{2-7}
$P4$&$P4_2$ & Ch & $\mathbb{Z}_2$& -- & 1 & 77\\\cline{2-7}
&$I4$ & Ch & $\mathbb{Z}_2$& -- & 1  &79\\\hline
$P3$&$P3$ & Ch & $\mathbb{Z}_3$& -- & 1 & 143, 173\\\hline
$P6$&$P6$ & Ch & $\mathbb{Z}_6$& -- &1 &168\\\hline\hline
& &  & $(\mathbb{Z}_2)^2$ & & $\mathbb{Z}_2$ & 10\\
$P2/m$&$P2/m$ & mCh & $\mathbb{Z}_2$ & mCh & $\mathbb{Z}_2$ & 51, 55\\
& &  & 1 & & $\mathbb{Z}_2$ & 49, 53, 58, 66\\\hline
& &  & $(\mathbb{Z}_4)^2$ & & $\mathbb{Z}_4$ & 83\\
&$P4/m$ & mCh & $\mathbb{Z}_4$ & mCh & $\mathbb{Z}_4$ & 127\\
& & & $1$ & & $\mathbb{Z}_4$& 124, 128\\\cline{2-7}
$P4/m$&\multirow{2}{*}{$P4_2/m$}  & \multirow{2}{*}{mCh}  & $\mathbb{Z}_4$ & \multirow{2}{*}{mCh} & $\mathbb{Z}_2$ & 84\\
& &  & 1 & & $\mathbb{Z}_4$ & 132, 135, 136, 223\\\cline{2-7}
&  \multirow{2}{*}{$I4/m$}  & \multirow{2}{*}{mCh}   & $\mathbb{Z}_4$ & \multirow{2}{*}{mCh} & $\mathbb{Z}_4$ & 87\\
& & & 1 & & $\mathbb{Z}_4$ & 140\\\hline
&   &  & $(\mathbb{Z}_3)^2$ & & $\mathbb{Z}_3$& 174\\
$P\bar{6}$&$P\bar{6}$ & mCh & $\mathbb{Z}_3$ & mCh &  $\mathbb{Z}_3$& 187, 189\\
& &  & $1$ & & $\mathbb{Z}_3$& 188, 190\\\hline
& \multirow{2}{*}{$P6/m$} &  \multirow{2}{*}{mCh}  & $(\mathbb{Z}_6)^2$ &  \multirow{2}{*}{mCh} & $\mathbb{Z}_6$& 175\\
\multirow{2}{*}{$P6/m$}& &  & $1$ & & $\mathbb{Z}_6$& 192 \\\cline{2-7}
 & \multirow{2}{*}{$P6_3/m$} &  \multirow{2}{*}{mCh}   & $\mathbb{Z}_3$&  \multirow{2}{*}{mCh}  & $\mathbb{Z}_6$& 176\\
&&    & 1 & & $\mathbb{Z}_6$& 193, 194\\ \hline\hline
&  &    & $\mathbb{Z}_4$ & Ch, $\mu_4$ & $(\mathbb{Z}_2)^2$& 81 \\
& & & $\mathbb{Z}_4$ & $\mu_4$ & $\mathbb{Z}_2$& 85 \\
& $P\bar{4}$ & Ch & $\mathbb{Z}_2$ & $\mu_4$ & $\mathbb{Z}_2$& 86 \\
$P\bar{4}$& &   & \multirow{2}{*}{$1$} & \multirow{2}{*}{$\mu_4$} & \multirow{2}{*}{$\mathbb{Z}_2$}& 111--118, 125, 126, 129, 130, 133, \\
& &  & & & &  134, 137, 138, 215, 218, 222, 224\\\cline{2-7}
&\multirow{2}{*}{$I\bar{4}$} & \multirow{2}{*}{Ch}  & $\mathbb{Z}_2$ & Ch, $\mu_4$ & $(\mathbb{Z}_2)^2$& 82\\
& &  & $1$ & $\mu_4$ & $\mathbb{Z}_2$& 119--122, 141, 142, 216, 217, 219, 220, 227, 228, 230\\\hline\hline
&  & & $(\mathbb{Z}_2)^3$ & & $\mathbb{Z}_4$ & 47\\
& $Pmmm$ & mCh  & $\mathbb{Z}_2$ &  $\kappa_1$ & $\mathbb{Z}_4$ & 131, 200\\
& & & 1 & & $\mathbb{Z}_4$ & 223\\\cline{2-7}
$Pmmm$&   \multirow{2}{*}{$Cmmm$} &    \multirow{2}{*}{mCh}  & $\mathbb{Z}_2$ &  \multirow{2}{*}{$\kappa_1$} & $\mathbb{Z}_4$ & 65\\
& &   & 1 & & $\mathbb{Z}_4$ & 132, 136\\\cline{2-7}
&$Fmmm$ & -- & 1 & $\kappa_1$ & $\mathbb{Z}_4$ & 69, 140, 202, 226\\\cline{2-7}
&$Immm$ & -- & 1 & $\kappa_1$ & $\mathbb{Z}_4$ &  71, 204\\\hline\hline
&\multirow{2}{*}{$P4/mmm$}  &\multirow{2}{*}{mCh} & $\mathbb{Z}_2\times\mathbb{Z}_4$ & \multirow{2}{*}{$\kappa_1-2\kappa_4$} &  $\mathbb{Z}_{8}$& 123\\
$P4/mmm$& &  & $\mathbb{Z}_4$ & & $\mathbb{Z}_{8}$& 221\\\cline{2-7}
& $I4/mmm$ & --  & 1& $\kappa_1-2\kappa_4$& $\mathbb{Z}_{8}$& 139, 225, 229\\\hline\hline
$P\bar{1}$, $P\bar{4}$ &   $I4_1/a$ & -- & 1 & $\mu_1$, $\mu_4$ & $(\mathbb{Z}_2)^2$& 88\\\hline
$P\bar{1}$, $P3 $& $P\bar{3}$ & Ch  & $\mathbb{Z}_2$ & Ch, $\mu_1$ & $\mathbb{Z}_{12}$& 147\\\hline
$Pmmm$, $P6/m$&$P6/mmm$ & mCh  & $\mathbb{Z}_6$ & mCh, $\kappa_1$ & $\mathbb{Z}_{12}$& 191\\\hline\hline
\end{tabular}
\end{center}
\end{table*}

\begin{table*}
\begin{center}
\caption{\label{tab:app2}\textbf{Symmetry indicators for spinless electrons in class A for 230 space groups.} }
\begin{tabular}{c|c|cc|cc|c}\hline \hline
Key SG & $G_0$ & weak & $X_{\textbf{BS}}^{(w)}$ & strong & $X_{\textbf{BS}}^{(s)}$ & Supergroups \\\hline
& & & $(\mathbb{Z}_2)^3$ & & $\mathbb{Z}_4$& 2\\
& &  & $\mathbb{Z}_2$ & & $\mathbb{Z}_4$& 148\\
$P\bar{1}$&$P\bar{1}$ & Ch & $\mathbb{Z}_2$ & $\mu_1$ & $\mathbb{Z}_2$& 12, 13, 15\\
& &  & \multirow{2}{*}{1}  &  & \multirow{2}{*}{$\mathbb{Z}_2$} & 11, 14, 48--50, 52--54, 56, 58, 60, 61, 66, 68, 70, 73, 126, \\ 
& &  & & & & 130, 133, 142, 162--167, 201, 203, 205, 206, 222, 228, 230\\\hline\hline
$P2$ &$P2$ & Ch & $\mathbb{Z}_2$& -- &1 & 3, 171, 172\\\hline
&$P4$ & Ch & $\mathbb{Z}_4$& -- &1  & 75\\\cline{2-7}
$P4$&$P4_2$ & Ch & $\mathbb{Z}_2$& -- & 1 & 77\\\cline{2-7}
&$I4$ & Ch & $\mathbb{Z}_2$& -- & 1  &79\\\hline
$P3$&$P3$ & Ch & $\mathbb{Z}_3$& -- & 1 & 143, 173\\\hline
$P6$&$P6$ & Ch & $\mathbb{Z}_6$& -- &1 &168\\\hline\hline
\multirow{2}{*}{$P2/m$} &\multirow{2}{*}{$P2/m$} &  \multirow{2}{*}{mCh}  & $(\mathbb{Z}_2)^2$ &  \multirow{2}{*}{mCh} & $\mathbb{Z}_2$ & 10\\
& &  & 1 &  & $\mathbb{Z}_2$ & 49, 53, 58, 66\\\hline
&  \multirow{2}{*}{$P4/m$}&  \multirow{2}{*}{mCh} & $(\mathbb{Z}_4)^2$ &  \multirow{2}{*}{mCh} & $\mathbb{Z}_4$ & 83\\
\multirow{2}{*}{$P4/m$}& & & 1 & & $\mathbb{Z}_4$ & 124, 128\\\cline{2-7}
&$P4_2/m$  & mCh  & $\mathbb{Z}_4$ & mCh & $\mathbb{Z}_2$ & 84\\\cline{2-7}
 &$I4/m$  & mCh   & $\mathbb{Z}_4$ & mCh & $\mathbb{Z}_4$ & 87\\\hline
\multirow{2}{*}{$P\bar{6}$}   &\multirow{2}{*}{$P\bar{6}$} &   \multirow{2}{*}{mCh} & $(\mathbb{Z}_3)^2$ &  \multirow{2}{*}{mCh} & $\mathbb{Z}_3$& 174\\
 &&  &  $1$ &  & $\mathbb{Z}_3$& 188, 190\\\hline
 &\multirow{2}{*}{$P6/m$} &  \multirow{2}{*}{mCh}  & $(\mathbb{Z}_6)^2$ &  \multirow{2}{*}{mCh} & $\mathbb{Z}_6$& 175\\
$P6/m$ &&  & $1$ & & $\mathbb{Z}_6$& 192 \\\cline{2-7}
&$P6_3/m$ &  mCh   & $\mathbb{Z}_3$& mCh  & $\mathbb{Z}_6$& 176\\\hline\hline
 &\multirow{5}{*}{$P\bar{4}$} &   \multirow{5}{*}{Ch} & $\mathbb{Z}_4$ & Ch, $\mu_4$ & $(\mathbb{Z}_2)^2$& 81 \\
 && & $\mathbb{Z}_4$ & $\mu_4$ & $\mathbb{Z}_2$& 85 \\
\multirow{2}{*}{$P\bar{4}$} && & $\mathbb{Z}_2$ & $\mu_4$ & $\mathbb{Z}_2$& 86 \\
 &&   & $1$ & $\mu_4$ & $\mathbb{Z}_2$&  112, 114, 116, 117, 118, 126, 130, 133, 218, 222\\\cline{2-7}
&\multirow{2}{*}{$I\bar{4}$} & \multirow{2}{*}{Ch}  & $\mathbb{Z}_2$ & Ch, $\mu_4$ & $(\mathbb{Z}_2)^2$& 82\\
 &&  & $1$ & $\mu_4$ & $\mathbb{Z}_2$& 120, 122, 142, 219, 220, 228, 230\\\hline\hline
$P\bar{1}$, $P\bar{4}$  &$I4_1/a$ & -- & 1 & $\mu_1$, $\mu_4$ & $\mathbb{Z}_2\times\mathbb{Z}_{2}$& 88\\\hline
 $P\bar{1}$, $P3$    &$P\bar{3}$ & Ch  & $\mathbb{Z}_2$ & Ch, $\mu_1$ & $\mathbb{Z}_{12}$& 147\\\hline\hline
&$Pcc2$  & -- & 1 & $\mu_2$ & $\mathbb{Z}_2$& 27, 49, 54, 56, 103, 116, 130\\\cline{2-7}
$Pcc2$&$Ccc2$  & -- & 1 &  $\mu_2$ & $\mathbb{Z}_2$&37, 66, 103, 112, 130, 184\\\cline{2-7}
&$Iba2$  & -- & 1&  $\mu_2$ & $\mathbb{Z}_2$& 45, 120, 142\\  \hline\hline
$P4nc$  &$P4nc$  & -- & 1&  $\nu$, $\mu_2$ & $\mathbb{Z}_2$&104 \\ \hline
 $P4_2bc$ &$P4_2bc$ & -- & 1&  $\nu$  &$\mathbb{Z}_2$& 106\\ \hline
$I4_1cd$&$I4_1cd$ & -- & 1 &  $\nu$  &$\mathbb{Z}_2$&110\\ \hline\hline
\end{tabular}
\end{center}
\end{table*}

\end{document}